# Analysis of gamma-amino butyric acid in the germinated brown rice by pre-column derivatization with 2, 4-dinitrofluorobenzene

**Hyok-Chol Ri**[1], **Jong-Ho Yun**[1], **Kye-Ryong Sin**[2*], **Su-Ryon Ri**[1]

(1: Institute of analysis, **Kim Il Sung** University, Pyongyang, DPR of Korea,
2: Faculty of chemistry, **Kim Il Sung** University, Pyongyang, DPR of Korea)
* E-mail: ryongnam9@yahoo.com

**Abatract:**

This paper reports the analytical method for gamma-amino butyric acid (GABA) in the germinated brown rice by liquid chromatography. GABA was derivatized by pre-column derivatization with 2, 4-dinitrofluorobenzene (DNFB). To separate GABA in the germinated brown rice extracts containing various amino acids, we reviewed the influence of the kind of separation columns and the gradient of mobile phase of ultra performance liquid chromatography (UPLC). GABA was extracted with distilled water from the germinated brown rice with the optimum extraction time of 12 h. To validate this method the precision and the recovery were discussed. In the concentration range from 5 to 50μg/mL, the calibration curve for GABA was linear and the regression equation was obtained with correlation coefficient ($R^2$) of 0.999. GABA was analyzed by amino acid analyzer in comparison with this method and the results of F-test and t-test indicated that there were no significant differences in accuracy and precision between two methods.

## 1. Introduction

Rice is the most staple and popular cereal food for about half of the world population in many countries. For populations living in many developing countries, rice contributes the greatest percentage of calories and protein. Rice consumers eat white rice and consider brown rice for the poor and animals. But now scientists have found out that brown rice is far better and healthier than white rice (Vetha et al. 2013). Brown rice contains more nutritional components such as dietary fibre, oryzanol, vitamin E, vitamin B and gamma-aminobutyric acid (GABA) than ordinary white rice. These components exist in the germ and bran layers of rice, most of which are removed by polishing or milling because brown rice is harder to chew and not as tasty as white rice (Champagne et al. 2004). The germinated brown rice gets sweeter, softer, swelled and cohesive than the ungerminated one (Thitima et al. 2012).

Germinated brown rice (GBR) is a functional food as it contains GABA which promotes brain health (Swati et al. 2011; Thitima et al. 2012). GABA is a non-protein amino acid and has been reported to provide beneficial effects for human health (Wichamanee et al. 2012). It can lower hypertension, promote sleepiness, reduce autonomic disorder observed during the menopausal period and inhibit cancer cell proliferation (Mustapha et al. 2012). Many studies have reported that GBR has more contents of GABA and vitamin B than the ungerminated brown rice and white rice (Anuchita et al. 2010; Rungtip et al. 2012; Trachoo et al. 2006).

There are many methods for analysis of GABA but the methods by UPLC or by amino acid analyzer have been mainly used (Anthony et al. 2014; Anuchita et al. 2010; Hannelore et al. 2009; Lin et al. 1980; Mustafa et al. 2007; Nurullah et al. 2015; Zhang et al. 2014). In the UPLC methods the derivatization reagents such as PITC (phenylisothiocyanate), FMOC-Cl (9–fluorenyl methyl chlorformate), Dansyl-Cl, AQC (6–aminoquinolyl–N–hydroxy–succinimidyl carbamate) and DNFB (2, 4-dinitrofluorobenzene) are used. Among those, the most widely used one is OPA (*o*-phtaldialdehyde), which offers the short derivatization time and simple sample preparation is very simple, but OPA amino acid derivatives are not stable. In case of DNFB, the dervatization time is little long but sample preparation is simple and its derivatives are very stable.

In this paper, presented was a method of analyzing GABA in the GBR by UPLC with pre-column derivatization using DNFB.

## 2. Materials and Methods

### 2.1. Reagents and apparatus

GABA standard (>99%) was obtained from Sigma-Aldrich. Amino acids mixture standard solution (type H) was obtained from Wako Pure Chemical Industries, Ltd. HPLC-grade acetonitrile was obtained from Honeywell Burdick. All other reagents and standards were obtained from Sigma-Aldrich. The LC analysis was performed using an ACQUITY ultra performance liquid chromatography-PDA. An ACQUITY UPLC HSS T3(1.8 $\mu$m, 2.1×150mm) and an ACCQ-TAG ULTRA C18(1.7 $\mu$m, 2.1×100mm) were used at the flow rate of 0.3 mL·min$^{-1}$ and at 30℃. The Amino Acid Analyzer (L-8900 System: Hitachi Inc.), equipped with a visible detector, was used for analysis of GABA and the individual amino acids. The DC 8006 constant temp-bath and Allegra X-12 Centrifuge were used.

### 2.2. Sample preparation procedures

GBR was made by germinating two Korean rice cultivars (*SH7* and *PY49*). Each of rice samples was soaked in distilled water at room temperature for 5h. Next the rice grains were taken out and were germinated in closed vessel at 30℃ for 36h. After germination the rice samples were dried to the moisture content below 13% at 50℃.

## 2.3. Optimization of extraction time for GABA from the GBR

The ground GBR samples were soaked in distilled water and extracted at 30℃ for different times 3, 6, 12, 18hours. Then the extract was filtered and the filtrate was analyzed for GABA content using both UPLC and amino acid analyzer.

## 2.4. Derivatization procedure

One-half to one gram (0.5-1g) of ground GBR samples were weighed in plastic tubes. Then 5mL of distilled water was added and the mixtures were incubated in the constant temp-bath at 30℃ for 12 h. Thereafter, 5mL of 6% trichloroacetic acid was added and the mixtures were centrifuged at 5000 rpm for 10 min. Next 0.5mL of the supernatants was added to 0.5mL of 0.2M borate buffer (pH 9.0) and then 0.5mL of 0.1% dinitrofluorobenzene was added. And the mixture was incubated at 60℃ of the constant temp-bath for 1h. Then 3.5mL of 0.1M phosphate buffer (pH 7.0) was added and 10μL of this solution was injected into autosampler.

## 2.5. Chromatographic methods of UPLC-PDA

The derivatized samples (10 μL injection) were separated on a column using solvent gradient with Eluent Buffer A (0.2M actate buffer (pH 5.8) containing 1% N, N'-DMF) and Eluent B (50:50(v:v) acetonitrile:ultrapure water solvent). The solvent gradient programs and buffer composition used are described in Table 1. The derivatized amino acids were detected using PDA detector (Waters) at 360nm and run-time was 13min.

Table 1. The solvent gradient program

| Time (min) | Eluent buffer A (%) | Eluent B (%) |
|---|---|---|
| 0 | 86 | 14 |
| 0.2 | 86 | 14 |
| 3 | 69 | 31 |
| 5 | 64 | 36 |
| 9 | 20 | 80 |
| 11 | 86 | 14 |
| 13 | 86 | 14 |

## 2.6. Chromatographic methods of amino acid analyzer

Amino Acid Analyzer (L-8900 System: Hitachi Inc.), equipped with a visible detector, was used for GABA analysis. Ion Exchange Column # 2622SC-PH (3 ㎛, 4.6mm×60mm) and guard 2650# (4.6 mm×40mm) columns were used. After injection into the columns, the auto-sampler was used for the inline-derivatization by Ninhydrin post-column derivatization. The Ninhydrin-derivatized amino acids were detected at 570 nm and at 440 nm.

## 3. Results and discussion
### 3.1. Optimization of separation condition of DNP-GABA by UPLC

Fig. 1 showed the chromatograms of different standards derivatized with DNFB under the gradient condition such as Table 1, where (A) showed the chromatogram of GABA derivatized with DNFB. As indicated in (A) the retention time of GABA was 2.70min. Fig.1(B) showed the chroamtogram of 17 kinds of amino acid standards derivatized with DNFB. Fig.1(C) showed the chromatogram of mixture of 17 amino acids and GABA. As indicated in Fig.1(B) and Fig.1(C) the peak of DNP-GABA was separated from peaks of 17 kinds of amino acids completely and the retention time of DNP-GABA was 2.69min. Therefore under the gradient condition on Table 1, it was confirmed that GABA was separated completely from the other amino acids.

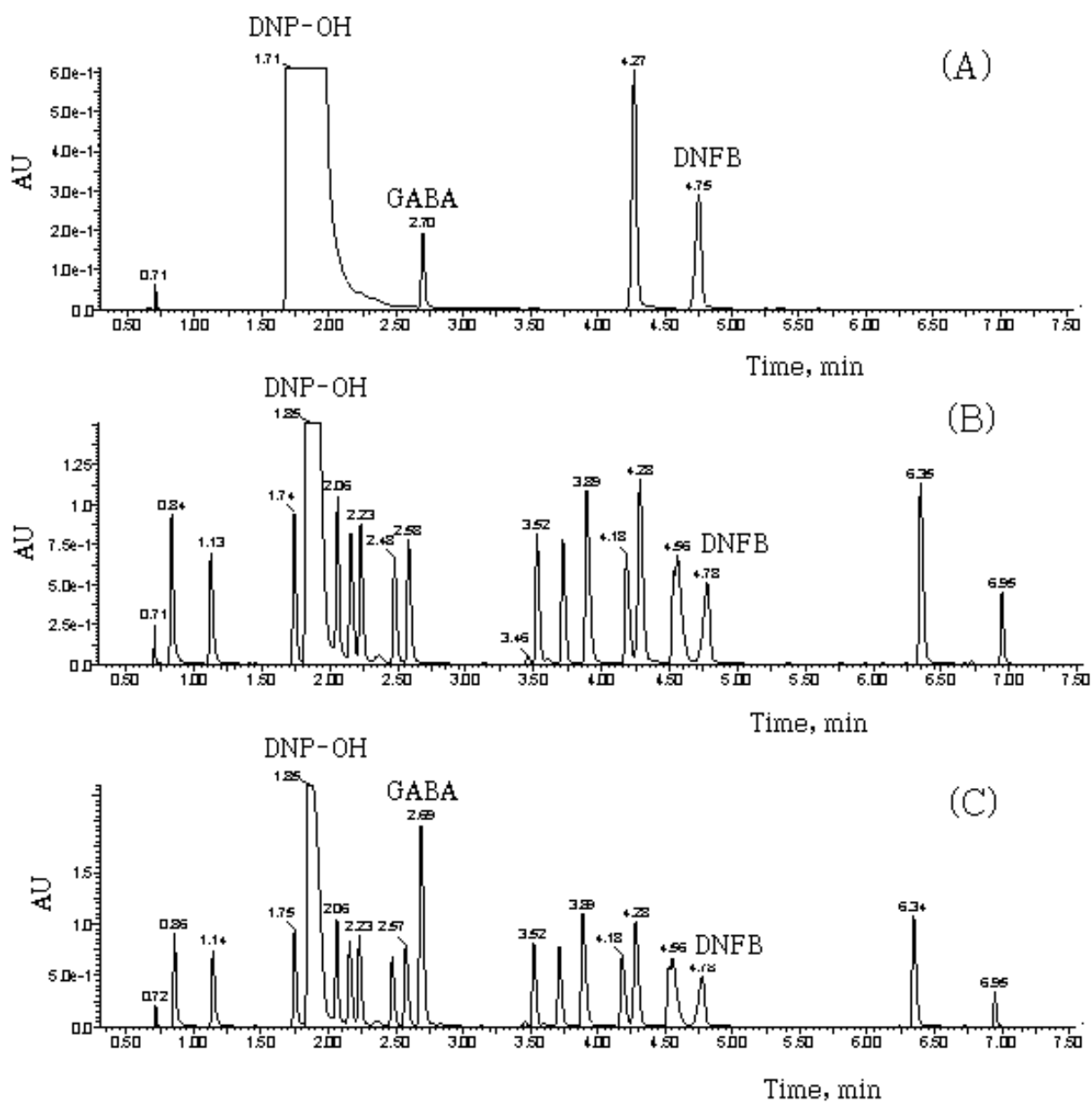

Figure 1. Chomatograms of DNP-GABA standard (A), amino acids mixture standards derivatized with DNFB (B) and amino acids mixture standards including GABA derivatized with DNFB (C)

Fig. 2 showed the chromatograms of DNP-GABA using the HSS T3 column and ACCQ-TAG ULTRA C18 column.

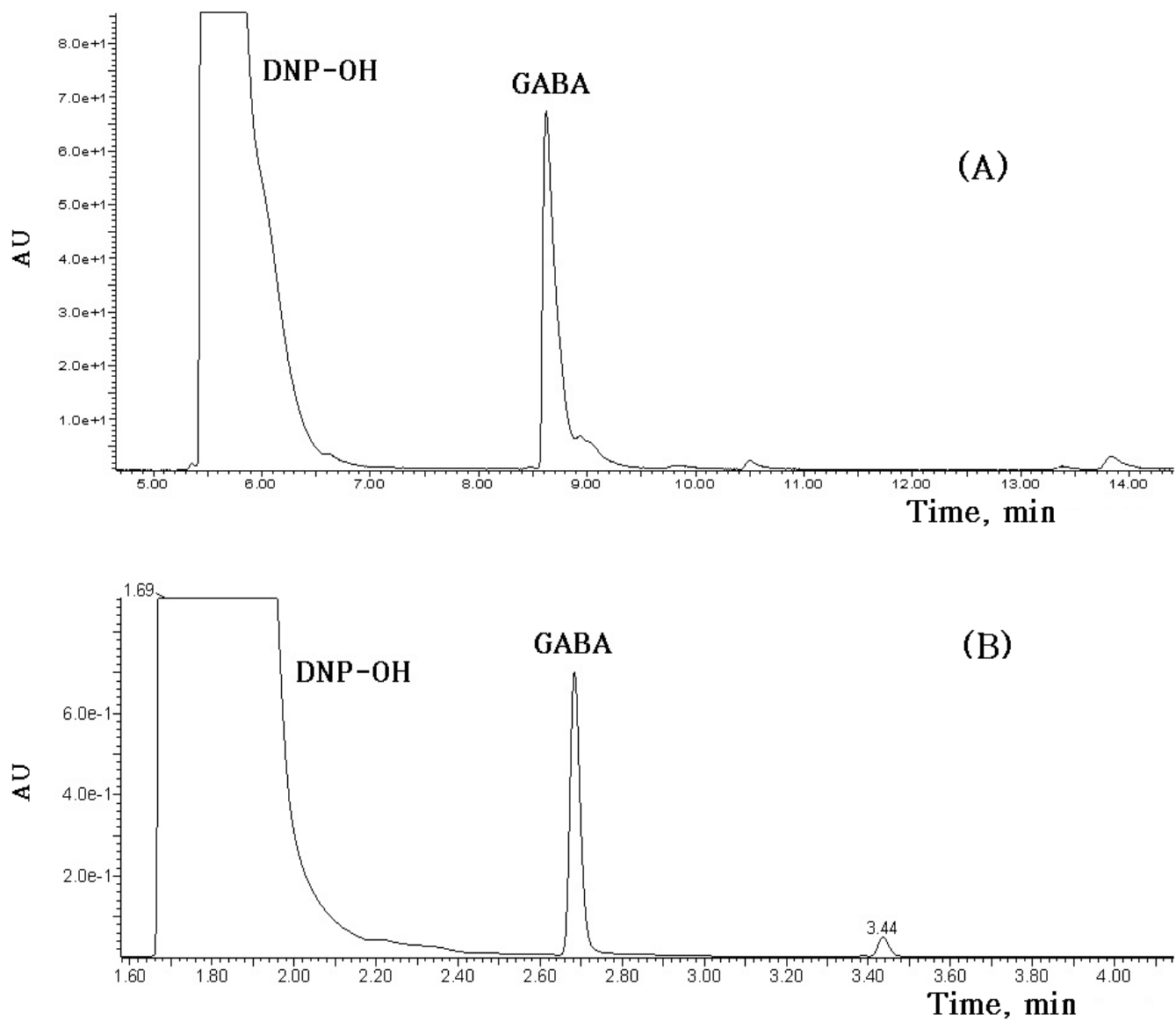

Figure 2. Chromatograms of DNP-GABA using the HSS T3 column (A) and
the ACCQ-TAG ULTRA C18 column (B)

Fig. 3 showed the chromatograms of 17 kinds of amino acids including GABA using the HSS T3 column and ACCQ-TAG ULTRA C18 column.

As indicated in Fig. 2 (A) and Fig. 3 (A) the peaks of all amino acids including GABA were tailed using the HSS T3 column, but peak tailing was not shown using ACCQ-TAG ULTRA C18 column. Therefore in the present study ACCQ-TAG ULTRA C18 column was selected.

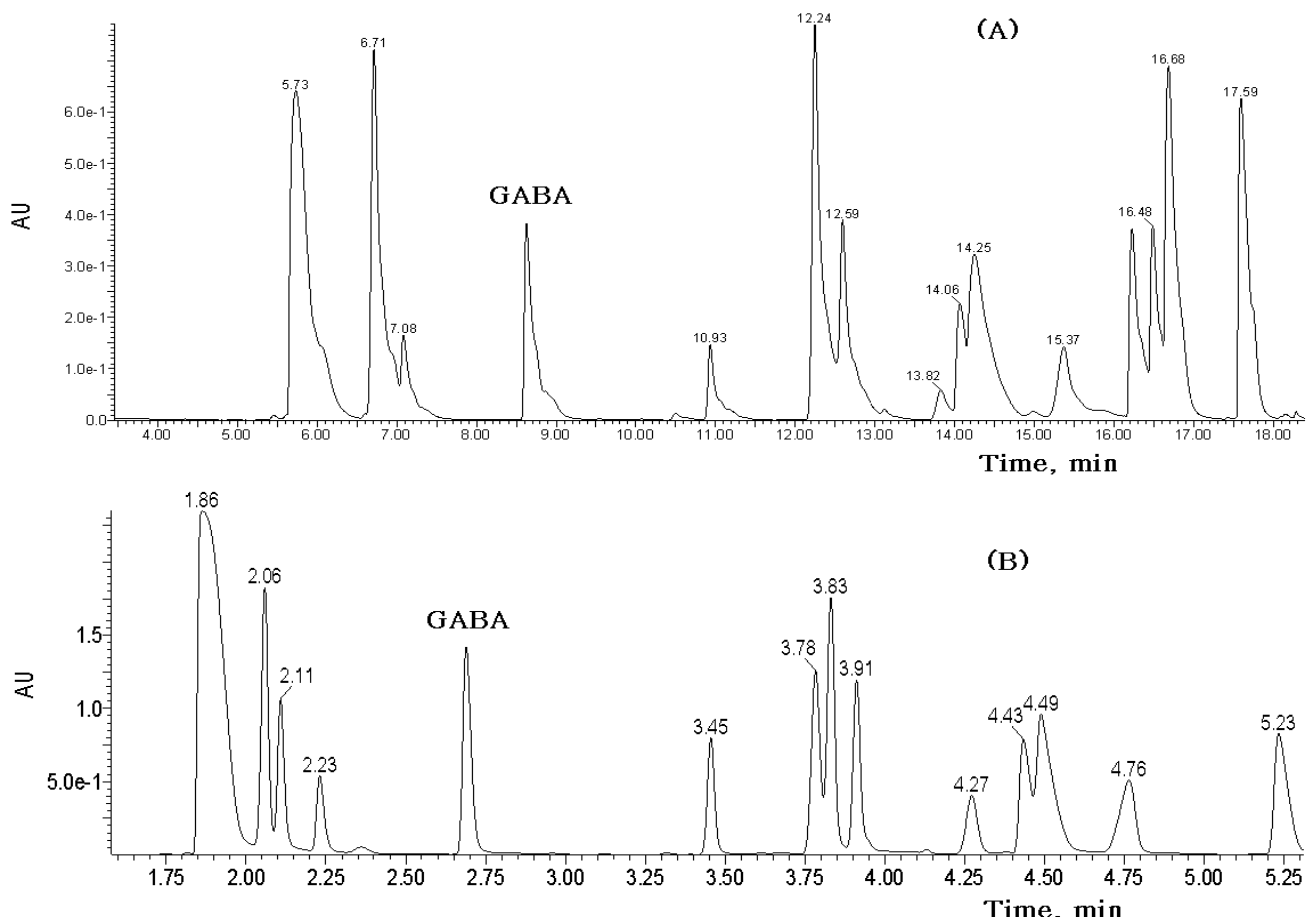

Figure 3. Chromatograms of 17 kinds of amino acids including GABA using the HSS T3 column (A) and the ACCQ-TAG ULTRA C18 column (B)

## 3.2. Method validation

### 3.2.1. Calibration curve for GABA using UPLC

The calibration curve for GABA was made by analyzing a series of GABA standard solutions in the concentration range from 5 to 50μg/mL. As indicated in Fig. 4 the calibration curve for GABA was linear and the regression equation was y=93.16x+341.6 with correlation coefficient ($R^2$) of 0.999.

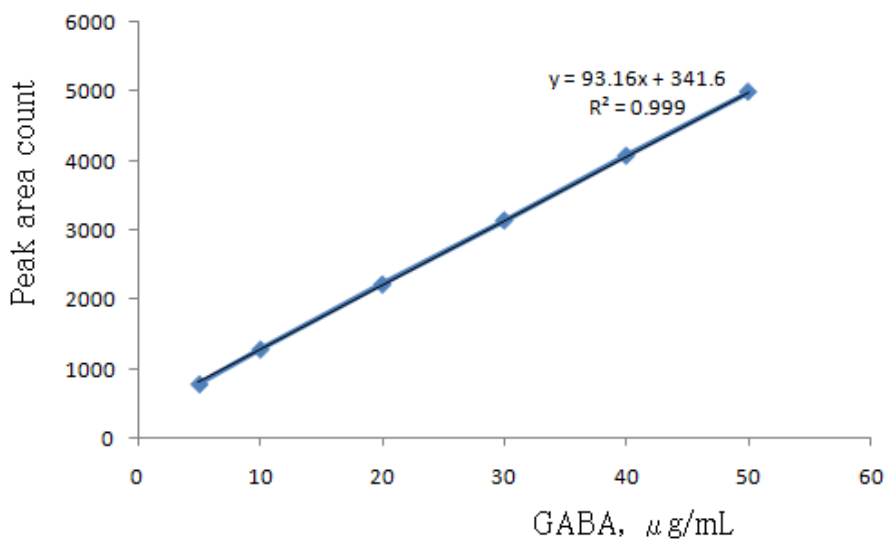

Figure 4.  Typical calibration for GABA

### 3.2.2. Precision

The intra-day and inter-day precision were obtained by analyzing the standard samples in the concentration range from 10 to 50μg/mL in three replicates within 1 day and on 3 days. Table 2 showed that the intra-day RSD were 1.28%, 0.95% and 1.12%, respectively, and the inter-day RSD were 1.05%, 1.43% and 1.25%.

Table 2.  Precision for the quantification of GABA

| Added Concentration (μg/mL) | Found Concentration (μg/mL) | Intra-day RSD(%) | Inter-day RSD(%) |
|---|---|---|---|
| 10 | 10.3 | 1.28 | 1.05 |
| 20 | 20.7 | 0.95 | 1.43 |
| 50 | 51.1 | 1.12 | 1.25 |

### 3.2.3. Recovery

The recovery was tested using the standard addition method. Proper amount of GABA was added to 0.5mL of GBR extract with known GABA content and the mixture was processed and analyzed as described in section 2.4 and 2.5. The recovery of GABA was 99.1% (n=3).

### 3.3. Optimization of GABA extraction time

Suitable amounts of distilled water were added to the ground GBR (two Korean rice cultivars *PY49* and *SH7*) and GABA was extracted at 30℃ for 3, 6, 9, 12, 15 and 18 h, respectively. As indicated in Fig. 5 the highest GABA content in the GBR was obtained when the GBR was extracted for 12h.

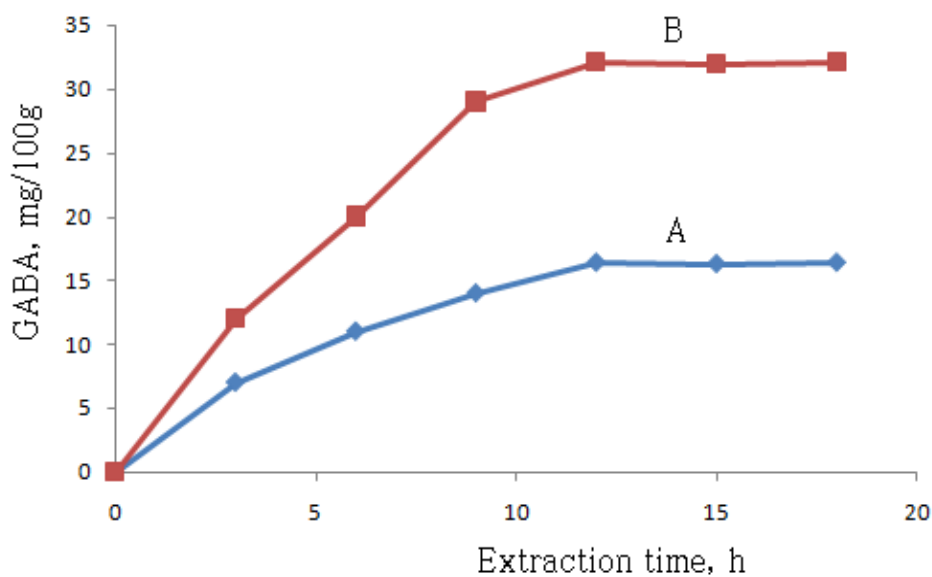

Figure 5. GABA content in the GBR extracts at various extraction times

(A: Korean rice cultivar *PY49*,  B: Korean rice cultivar *SH7*)

**3.4. Analysis for GABA in the germinated Korean rice cultivars**

Table 3 showed the results of analysis using UPLC and amino acid analyzer (AAA) for GABA in the GBR, which were germinated from Korean rice cultivars.

Table 3. Results of analysis for GABA in the GBR of Korean rice cultivars (mg/100g)

| Samples | Analyzer | 1 | 2 | 3 | 4 | Average | SD | RSD (%) |
|---|---|---|---|---|---|---|---|---|
| *SH7* | UPLC | 32.5 | 31.7 | 32.1 | 31.9 | 32.1 | 0.36 | 1.14 |
| | AAA | 31.8 | 32.2 | 31.9 | 32.1 | 32.0 | 0.18 | 0.57 |
| *PY49* | UPLC | 16.5 | 16.3 | 16.8 | 15.9 | 16.4 | 0.38 | 2.30 |
| | AAA | 15.9 | 16.0 | 15.8 | 16.2 | 16.0 | 0.17 | 1.07 |

As shown in Table 3, the content of GABA in *SH7* was 32.1±0.4mg/100g and the content of GABA in *PY49* was 16.4±0.4mg/100g by using UPLC. To evaluate the precision and accuracy of this method, the results from UPLC analysis were compared with that of AAA. According to F-test ad t-test, F value was 2 and it was lower than F(3, 3, 0.05), and t value was 0.8607 and it was lower than t(3, 3, 0.05). It means that there were no significant differences of precision and accuracy between two methods. The advantage of UPLC is that the run-time was very short compared with AAA.

**4. Conclusion**

In this paper, the analytical method for GABA in the GBR by UPLC was presented and validated. The extraction conditions of GABA from two GBR species (*SH7* and *PY49*) cultivated in Democratic People's Republic of Korea were found and GABA content was determined. GABA contents of two kinds of the GBR were 32.1±0.4mg/100g and 16.4±0.4mg/100g, respectively. The accuracy, precision and recovery of separation and quantification of GABA by this method were satisfactory. There were no significant differences of precision and accuracy between the UPLC method and AAA method.

**Compliance with Ethical Standards**

Conflict of Interest: Every author (A, B, C, D) declares that he has no conflict of interest.

Ethical approval: This article does not contain any studies with human participants or animals performed by any of the authors.